\documentclass[aps,prb,showpacs,twocolumn]{revtex4}
\usepackage{color}
\usepackage[utf8x]{inputenc}
\usepackage{amssymb}
\usepackage{graphicx}
\newcommand{\ud}{\mathrm{d}}

\begin{document}

\title{Paramagnetic Effects in the Vortex Lattice Field Distribution of Strongly Type-II Superconductors}

\author{V. P. Michal and V. P. Mineev}
\affiliation{Commissariat \`a l'Energie Atomique,
INAC/SPSMS, 38054 Grenoble, France}
\date{\today} 

\begin{abstract}
We present an analysis of the magnetic field distribution in the Abrikosov lattice of high-$\kappa$ superconductors with singlet pairing in the case where the critical field is mainly determined by the Pauli limit and the superfluid currents partly come from the paramagnetic interaction of electron spins with the local magnetic field. The derivation is performed in the frame of the generalized Clem variational method which is valid not too close to the critical field and furthermore with the Abrikosov type theory in the vicinity of it.
The found  vortex lattice form factor increases with increasing field and then falls down at approach of the upper critical field where the superconducting state is suppressed.
\end{abstract}
\maketitle

\section{Introduction}

Recent small angle neutron scattering experiments performed on the heavy-fermion superconductor CeCoIn$_5$
have revealed an unexpected behavior of the vortex lattice (VL) form factor\cite{Bianchi, White,Esk} defined as the Fourier transform of the local magnetic field in the vortex lattice. The VL form factor of the type II superconductors is usually a decreasing function of the magnetic field.\cite{Clem}
On the contrary, the VL form factor in CeCoIn$_5$ was found to increase with increasing magnetic field and then to fall down at the approach of upper critical field. \cite{Bianchi, White}

CeCoIn$_5$ is a tetragonal, d-wave pairing superconductor with a large Ginzburg-Landau (GL) parameter (i.e. $\kappa=\lambda/\xi\gg1$, where $\lambda$ and $\xi$ are the two characteristic lengths of the GL theory), and with the highest critical temperature
($T_c=2.3\,K$) among all the heavy fermion compounds.\cite{Petrovic,Izawa,Aoki}
It  has already generated great interest caused by the signs of the existence of the FFLO phase for a magnetic field parallel to the ab plane (and possibly to the c-axis),\cite{Bianchi2} and by the presence of a temperature interval $0<T<T_0$ where the superconducting/normal phase transition is of the first order.\cite{Izawa,Bianchi2,Bianchi3,Miclea}  The temperature $T_0$ was found equal to $0.3 T_c$ for the field orientation parallel to the tetragonal axis.\cite{Bianchi3}

The distance between the vortices in the mixed state of type II superconductors decreases with increasing magnetic field. It diminishes with respect to field penetration depth. As a result the field distribution is smoothed towards $H_{c2}$. Hence the form factor characterizing the sharpness of local field modulation is also inevitably diminished with increasing field. The unusual field dependence 
of the form factor in CeCoIn$_5$ points out  that some other mechanism leads to the opposite process, namely to the increasing of fast field modulation with field growth (Fig. \ref{hlog}).
The anomalous magnetic field dependence of the flux lattice form factor  observed by neutron scattering has been attributed to the large paramagnetic depairing effect in CeCoIn$_5$. \cite{Ichioka,Machida} 
 
The relative weight of the orbital and the Zeeman mechanisms for suppression of the superconducting state by a magnetic field is quantified by the Maki parameter $\alpha_M=\sqrt{2}H_{c20}/H_p$, where $H_{c20}=\phi_0/2\pi\xi_0^2$ is the orbital critical field while $H_p=\Delta_0/\sqrt{2}\mu$ is the Pauli limiting field. $\phi_0\simeq 2.07\times 10^{-7}\,G\cdot cm^2$ is the flux quantum. The Maki parameter is expressed through the Fermi velocity $v_F=k_F/m^*$ and the critical temperature as $\alpha_M\approx T_c/mv_F^2$, hence it takes larger values in heavy fermionic compounds where effective mass of charge carriers is much larger than the bare electron mass $m^*\gg m$.
Unlike the majority of superconductors, the Pauli limiting field in CeCoIn$_5$ is smaller\cite{Bianchi3} than the orbital critical field by a factor of $> 3$. Hence, the Zeeman interaction plays an important role in the mixed state field and current distributions.

The role of the paramagnetic mechanism has been investigated  using numerical processing of the quasi-classical Eilenberger equations.\cite{Ichioka,Machida}  In the first paper\cite{Ichioka}  the authors could not reproduce the form factor increasing behavior towards $H_{c2}$ observed in CeCoIn$_5$. Then in the subsequent calculation \cite{Machida} performed with an even stronger paramagnetic contribution they have obtained the desirable correspondence. The developed numerical procedure has been  performed at temperature $T=0.2 T_c$ where the phase transition from the normal to the superconducting state is of the first order. In this  region of the phase diagram an analytical treatment of the problem is impossible. However, at temperatures higher than $T_0$ and at large Maki parameters one can develop an analytic theory giving the possibility to  compare directly the roles of the orbital and the Zeeman contributions in the form factor field dependence.

Our approach is based on the Clem \cite{Clem} elegant method which approximates the GL order parameter by a trial function and allows an analytic  solution of  the second GL equation for the local magnetic field. Here
we present an analytic derivation of the magnetic field distribution and VL form factor taking  the paramagnetic effects into account. 
For this purpose we shall use the GL theory developed in the papers\cite{Hou2,Hou1} where the diamagnetic superfluid currents are  determined not only by the orbital effects but also by the Zeeman interaction of the electron spins with the local magnetic field.
The Clem-type analysis which makes use of the isolated vortices approximation is valid not too close to the critical field. Near the critical field it should be completed by the magnetic field distribution based on the Abrikosov type distribution of the order parameter.  

By direct solution of the Maxwell equation including the currents due to electron spin interaction with 
magnetic field we demonstrate analytically that the VL form factor growth is related to sharp changes of the local magnetic field concentrated in the cores of vortices that is on a distance of the order of the coherence length around the vortex axis. 
The found form factor decreases with increasing magnetic field in the high temperature - low field  region 
of the phase diagram, while at lower temperatures it turns to the 
behavior increasing with increasing field
and then decreasing to zero at the approach of the critical field.  In accordance with numerical studies \cite{Ichioka,Machida} we have found that the non-monotonic behavior reveals itself in case of strong enough paramagnetic contribution.
The obtained results qualitatively describe the magnetic field dependence of the vortex lattice form factor in any high-$\kappa$ superconductor with singlet pairing and large Maki parameter. 
At the same time  we do not pretend on a quantitative correspondence of our findings with the CeCoIn$_5$ form factor dependence reported in the papers \cite{Bianchi, White,Esk}. In absence of knowledge of real band structure, the Fermi surface shape, pairing mechanism, probable field dependence of the effective mass, the value and the angular dependence of the $g$-factor, any theory including the developed numerical procedure \cite{Ichioka,Machida} cannot pretend on a quantitative correspondence with experiment. Therefore, our goal is to develop an analytical model which gives us a clear picture of the field distribution in the Abrikosov lattice of the superconductors with a large Maki parameter value.

The paper is organized as follows. In the Second Section we briefly repeat the Clem results and discuss the limitation of their applicability. Then, in the third Section the corresponding theory taking into account the Zeeman interaction is presented. The behavior of the form factor in the GL region near the critical field is described in Section IV. The Concluding remarks are formulated in Section V.

\section{Orbital form factor}

Regarding the form factor field dependence, the particular symmetry of the flux line lattice is unimportant. For simplicity we shall consider a square vortex lattice with axis spacing $a=\sqrt{\phi_0/B}$ formed in a type II superconductor under magnetic field directed along the z-axis. The magnetic induction $\mathbf{B}=\overline{\mathbf{h}}$ is determined as the spatial average of the local magnetic field
$\mathbf{h}=h\hat z=\nabla\times\mathbf{A}$. 
The form factor $F_{mn}$ is determined by the  Fourier transform of the magnetic field
\begin{equation}
 F_{mn}=h(\mathbf{q}_{mn})=\int\ud^2\mathbf{r}\,h(\mathbf{r})e^{-i\mathbf{q}_{mn}\cdot\mathbf{r}}
\end{equation}
where $\mathbf{q}_{mn}=\frac{2\pi}{a}(m\hat{x}+n\hat{y})$ are the vectors of the reciprocal lattice.  For the external field not too close to $H_{c2}$, where the distance between the vortices is larger than the core radius, the local magnetic field represents the sum of the magnetic fields of separate vortices $ h({\bf r})=\sum_i h_v({\bf r}-{\bf r}_i)$. Thus, the form factor is proportional to the Fourier transform of magnetic field around one vortex 
\begin{equation}
 F_{mn}=\frac{B}{\phi_0}h_v(\mathbf{q}_{mn})
 \end{equation}
 This quantity is related to the intensity of  Bragg peaks observed in small angle neutron scattering experiments. Throughout the article we shall consider the form factor $F_{01}$ that corresponds to the indices (0,1).  We designate it as $F$.

The GL theory for the VL form factor, valid in the limit $\kappa\gg1$   was developed  by J. Clem. \cite{Clem} 
Starting from the general form of the order parameter for an isolated vortex
\begin{equation}
 \Delta(\mathbf{r})=\Delta_\infty f(r) e^{-i\varphi}
 \label{OP}
\end{equation}
($\varphi$ is the azimuthal angle in the plane perpendicular to the vortex axis), he proposed to model $f(r)$ by the trial function
\begin{equation}
 f(r)=\frac{r}{R},
 \label{f}
\end{equation}
with $R=\sqrt{r^2+\xi_v^2}$. The variational parameter $\xi_v$ was constrained to minimize the vortex total energy and was found to be in the large $\kappa$ limit $\xi_v=\sqrt{2}\,\xi$, where $\xi$ is the coherence length. Clem has calculated the field distribution due to the orbital current and  obtained the form factor
\begin{equation}
F_{orb}=B\,\frac{K_1(Q \xi_v)}{Q\lambda K_1(\xi_v/\lambda)},
\label{CFF}
\end{equation}
where  $Q=\sqrt{q^2+\lambda^{-2}}$, $q=2\pi\sqrt{B/\phi_0}$, $K_1(x)$ is the modified Bessel function of first order,\cite{Abr} and $\lambda$ is the London penetration depth.
One can write an approximate form of it in the conditions $\kappa\gg1$ and $q\gg\lambda^{-1}$, 
\begin{equation}
F_{orb}\simeq \frac{\phi_0}{(2\pi\lambda)^2}q\xi_vK_1(q\xi_v).
\label{Fgamma}
\end{equation}
The found form factor slowly decreases with magnetic field.  While the product $q\xi_v$ is smaller than 1, that is not too close to the upper critical field, the form factor  linearly decreases with field.  The formal application of  Eq. (\ref{Fgamma})  up to $H\approx H_{c2}$, where $q\xi_v\approx \sqrt{2\pi}$, gives the exponential decrease of the form factor. In fact in the vicinity of $H_{c2}$ the approximation of independent vortices does not work. The proper calculation  should be done using  the Abrikosov \cite{Abrikosov} form of the order parameter (see Section IV)  that leads  to the vanishing of the form factor linearly with $(H_{c2}-H)$.

Equation (\ref{CFF}) is valid in the region of validity of the GL theory. The latter does work near the critical temperature $(T_c-T)/T_c\ll 1$ but not at low temperatures where  the gradient expansion of the free energy is inapplicable and one needs to take into account the higher order gradient terms. Despite of this fact equation (\ref{CFF}) is widely used in the discussions
of the form factor field dependence at  low temperatures $T\ll T_c$ (see for instance Ref. 3, 17).
The higher order  gradient terms  change the concrete coordinate dependence of the order parameter near the vortex axis but details of  this dependence have no significant influence on the field distribution in the vortex lattice.
Hence, in the strong type II superconductors even at low temperatures Eq. (\ref{CFF}) still gives the qualitatively correct description of the form factor magnetic field dependence for the fields not too close to $H_{c2}$.

\section{The Zeeman contribution}

The form factor given by equations (\ref{CFF}), (\ref{Fgamma}) is  reliable if one neglects the paramagnetic interaction of the electron spins with the magnetic field. The latter leads to two extra features that are existent in the case of a large enough Maki parameter.
First, the two characteristic lengths $\xi$ and $\lambda$ in the above expression prove to be slightly magnetic field dependent. Second, a new mechanism originating from the Zeeman interaction gives rise to a contribution to the diamagnetic screening \cite{Hou2}
in the high magnetic field region of the phase diagram. To find it we start with the Ginzburg-Landau formulation including the paramagnetic effects.

The superconductor CeCoIn$_5$ has pairing symmetry $d_{x^2-y^2}$,\cite{Vorontsov} with order parameter
\begin{equation}
 \Delta_{\mathbf{k}}(\mathbf{r})=\psi(\mathbf{\hat{k}})\Delta(\mathbf{r}),~~~~\psi(\mathbf{\hat{k}})=\sqrt{2}\cos(2\varphi).
\end{equation}
For $s$-wave superconducting state $\psi(\mathbf{\hat{k}})=1$.
The free energy of the system is given by the GL functional
\begin{eqnarray}
\nonumber\mathcal{F}=\int\ud^2\mathbf{r}\,\Big(\,\frac{\mathbf{h}^2}{8\pi}+\alpha|\Delta|^2+
\varepsilon(h_{z}-B)|\Delta|^2\\
+\beta|\Delta|^4+\gamma|\mathbf{D}\Delta|^2\Big),
\label{GL}
\end{eqnarray}
where $\mathbf{D}=-i\nabla+2e\mathbf{A}$ is the gauge-invariant gradient (from here we put $\hbar=c=1$), ${\bf h}=rot{\bf A}$ is the local internal magnetic field,
and 
the coefficients in the functional depend on both  temperature $T$ and induction $B$ determined by the spacial average $\bar{\bf h}\equiv{\bf B}=B\hat z$. In the clean limit they are\cite{Hou2,Hou1}
\begin{eqnarray}
 \alpha&=&N_0\big(\ln(T/T_c)+\Re\mathfrak{e}\Psi(w)- \Psi (1/2)\big),\nonumber\\
 \varepsilon&=&\frac{d\alpha}{d B}=\frac{N_0\mu}{2\pi T}\Im\mathfrak{m}\Psi'(w),\nonumber\\
 \beta&=&-\frac{N_0}{8(2\pi T)^2}\langle|\psi(\mathbf{\hat{k}})|^4\rangle\Re\mathfrak{e}\Psi^{(2)}(w),\nonumber\\
 \gamma&=&-\frac{N_0 v_F^2}{8(2\pi  T)^2}\langle|\psi(\mathbf{\hat{k}})|^2\hat k_x^2\rangle\Re\mathfrak{e}\Psi^{(2)}(w),\nonumber
\end{eqnarray}
where $\Psi(w)$
is the digamma function, $\Psi^{(m)}(w)$ are its derivatives called by the polygamma functions, \cite{Abr} and 
$$
 w=\frac{1}{2}-\frac{i\mu B}{2\pi  T}.
$$

We shall consider $s$- and $d$-wave superconducting states in a quasi-two-dimensional crystal with a nearly cylindrical Fermi surface. For the $d$-wave order parameter given above the averages over the Fermi surface are $\langle|\psi(\mathbf{\hat{k}})|^4\rangle=3/2$ and $\langle|\psi(\mathbf{\hat{k}})|^2\hat k_x^2\rangle$=1/2 while for $s$-wave superconductivity and the same Fermi surface the corresponding averages are 1 and 1/2.

In the paramagnetic limit, when the orbital effect is neglected 
 the transition from the normal to a superconducting state takes place at the critical field $B_c(T)$ defined by equation $\alpha(T,B)=0$.  Along this transition line, the coefficients
$\beta(T,B)$ and $\gamma(T,B)$, which are positive near $T_c$, become negative at $T<T^*\simeq0.56T_c$.
This defines the tricritical point $(T^*,B^*)$ of the phase diagram with $B^*=B_c(T^*)\simeq1.07T_c/\mu$. At the tricritical point, the sign change of the coefficient $\gamma$ signals an instability toward the FFLO state with spatial modulation of the order parameter $\Delta$, while
the sign change of the coefficient $\beta$ signals a change of the order of the normal to superconducting phase transition. A more elaborate treatment including the orbital effects and the higher order terms in the GL functional \cite{Hou1} results in the following effects: (i) the upper critical field is slightly reduced by value  of the order  $B_c(T)/\alpha_M$; (ii) the temperature where the change of the order of the transition occurs and the one where the FFLO state arises are decreased by values of the order  $T_c/\alpha_M$ with respect to $T^*$. Below we consider only the temperatures above $T^*$ where the $\beta$ and $\gamma$ coefficients are positive and it is not necessary to take the higher order gradient terms into account. 
 
In the case of a large Maki parameter the GL expansion of the free energy in powers of the order parameter and its gradients is valid near the critical field  which is mainly determined by the paramagnetic depairing effect.\cite{Hou2,Hou1} At smaller fields strictly speaking the higher order gradient terms should be included.
The situation is similar to the application of the Clem formula at low temperatures discussed in the previous Section.
The higher order  gradient terms  change the concrete coordinate dependence of the order parameter near the vortex axis but details of  this dependence have no significant influence on the field distribution in the vortex lattice.
Hence, in the strong type II superconductors even for the fields noticeably smaller than the critical field the calculation with a GL functional containing just the second order gradient term
gives the qualitatively correct description of the form factor magnetic field dependence.{\cite{f}

Following Clem's procedure we consider an isolated vortex 
with order parameter given by Eqs. (\ref{OP}), (\ref{f})  and amplitude
$ \Delta_\infty=\sqrt{|\alpha|/2\beta}$. 
The field distribution around a single vortex is determined by the Maxwell equation derived from the stationary condition of the GL functional with respect to the vector potential 
\begin{equation}
\frac{1}{4\pi}\nabla\times\mathbf{h}_v={\bf j}.
\label{Maxwell}
\end{equation}
The density of current 
\begin{equation}
{\bf j}={\bf j}_{orb}+{\bf j}_Z
\end{equation}
consists of two parts originating from two different terms in the GL functional.
The \emph{orbital} density of current  is
\begin{equation}
{\bf j}_{orb}=-8e^2\gamma\big [A_v(r)-\frac{\phi_0}{2\pi r}\big]|\Delta|^2\hat{\varphi} ,
 \label{jgamma}
 \end{equation}
while the \emph{Zeeman} current \cite{Hou2} is
\begin{equation}
{\bf j}_{Z}=\varepsilon\frac{d}{dr}|\Delta|^2\hat{\varphi}.
\label{ZCurrent}
\end{equation}
Here the vector potential has the form ${\bf A}_v({\bf r})=A_v(r)\hat\varphi$.
Hence, with the help $\mathbf{h}_v=\nabla\times{\bf A}_v$, we come to the equation that determines the vector potential $A_v(r)$
\begin{equation}
 \frac{d}{dr}\big(\frac{1}{r}\frac{d}{dr}(rA_v)\big)-\frac{f^2}{\lambda^2}A_v=-\frac{\phi_0f^2}{2\pi\lambda^2r}-4\pi\varepsilon\Delta_\infty^2\frac{df^2}{dr},
 \label{DE}
\end{equation}
where $\lambda=\sqrt{\beta/16\pi e^2\gamma|\alpha|}$ is the penetration depth. 

Let us introduce the auxiliary function
$$v_s(r)=\frac{\phi_0}{2\pi r}-A_v(r)$$ 
playing the role of superfluid velocity \cite{Abrikosov} and 
substitute this into Eq. (\ref{DE}). We obtain the differential equation with an inhomogeneous term of Zeeman origin
\begin{equation}
  \frac{d}{dr}\big(\frac{1}{r}\frac{d}{dr}(rv_s)\big)-\frac{f^2}{\lambda^2}v_s=4\pi\varepsilon\Delta_\infty^2\frac{df^2}{dr}.
  \label{veleq}
\end{equation}
The general solution for this equation 
\begin{equation}
v_s(r)=v_{s}^i(r)+v_s^h(r)
\end{equation}
consists of the sum of particular solution of the inhomogeneous equation (\ref{veleq}) and a solution of corresponding homogeneous equation.
The former is given by 
\begin{equation}
v_{s}^i(r)=-\frac{R}{r}K_1(R/\lambda) C(R/\lambda),
\end{equation}
where
\begin{equation}
C(z)=-\frac{8\pi\varepsilon\Delta_\infty^2\xi_v^2}{\lambda}\int_{\xi_v/\lambda}^z \frac{d z}{zK_1^2(z)}
\int^z\frac{K_1(z)}{z^2}dz,
\end{equation}
chosen such that $v_{s}^i(0)=0$ and $v_{s}^i(\infty)=0$. The latter condition is assured by letting
the constant be zero in the primitive  $\int^z\frac{K_1(z)}{z^2}dz$ of the function $\frac{K_1(z)}{z^2}$.

The falling to zero at $r\rightarrow\infty$ solution of the homogeneous equation
\begin{equation}
v_s^h(r)=\frac{\phi_0}{2\pi \xi_v}\frac{RK_1(R/\lambda)}{rK_1(\xi_v/\lambda)}
\end{equation}
meets the requirement that the vector potential 
\begin{equation}
A_v(r)=\frac{\phi_0}{2\pi r}-v_s^h(r)-v_s^i(r)
\label{Aorb}
\end{equation}
vanishes on the vortex axis.
  
One can divide the total vector potential by its orbital part and its Zeeman part
\begin{equation}
A_v(r)=A_{orb}(r)+A_Z(r),
\label{Av}
\end{equation}
where the orbital part 
\begin{equation}
A_{orb}(r)=\frac{\phi_0}{2\pi r}-v_s^h(r)=\frac{\phi_0}{2\pi r}\left (1- \frac{RK_1(R/\lambda)}{\xi_vK_1(\xi_v/\lambda)}\right )
\end{equation}
is the solution of the equation (\ref{DE})
in the absence of the last term of the Zeeman origin that was found in Ref. 4. 
The corresponding magnetic field ${\bf h}_{orb}=h_{orb}\hat z$ is 
\begin{equation}
 h_{orb}=\frac{\phi_0}{2\pi\lambda\xi_v}\frac{K_0(R/\lambda)}{K_1(\xi_v/\lambda)},
\label{horb}
\end{equation}
and the form factor is determined by Eq. (\ref{CFF}).

The Zeeman part of the vector potential is given by
 \begin{equation}
 A_Z(r)=-v_s^i(r)=\frac{R}{r}K_1(R/\lambda) C(R/\lambda).
 \label{ZPotential}
 \end{equation}
 The corresponding magnetic field ${\bf h}_Z=h_Z\hat z$ reads
\begin{equation}
 h_Z(r)=\frac{1}{\lambda}\left[-K_0(R/\lambda)C(R/\lambda)+K_1(R/\lambda)C^\prime(R/\lambda)\right].
\label{heps}
\end{equation}
The numerically found magnetic field coordinate dependence is shown in Fig. 1.
\begin{figure}[b]
\centering
\includegraphics[width=9cm]{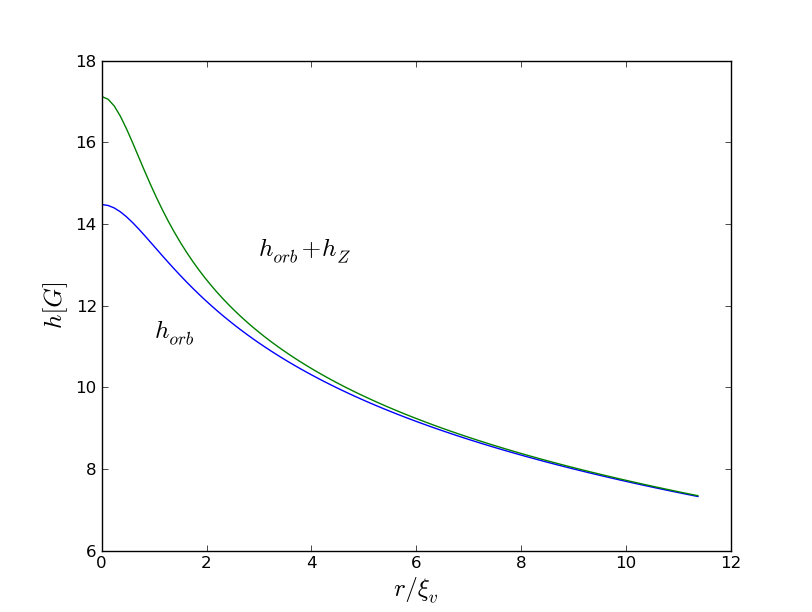}
\caption{Plot of the radial distribution of the magnetic field around a single vortex at induction $B=1.0\,T$ and temperature $T=1.05T^{\ast}$ calculated using parameters pointed out in text. The blue curve represents the orbital part of magnetic field (see the text)
 while the green one corresponds to the total field. The field $h_Z(r)$ has significant variation near the vortex core. While at large distances of the order of inter vortex spacing (not shown here) it starts to be negative, such that the total flux around single vortex is still equal to flux quantum (see text).}
\label{hlog}
\end{figure}

It is worth noting that the Zeeman term does not spoil the basic properties of the Abrikosov vortex.
Namely the total magnetic flux through the surface perpendicular to the vortex axis is equal to the flux quantum
\begin{eqnarray}
2\pi\int_0^\infty (h_{orb}(r)+h_Z(r))rdr=~~~~~~~~~~~~~~~~~~~~~~~~~~\nonumber\\
2\pi \lim_{r \rightarrow \infty} r(A_{orb}(r)+A_Z(r))=\phi_0~~~ 
\end{eqnarray} 
To prove this property, one must take into account the asymptotic behavior at large distances of functions 
\begin{equation}
K_1(z)\approx\sqrt{\frac{\pi}{2z}}e^{-z},~~~~~~~z\gg 1,
\end{equation} 
and 
\begin{equation}
C(z)\approx\frac{8\sqrt{2\pi} \varepsilon\Delta_\infty^2\xi_v^2}{\lambda } \frac{e^{z}}{z^{5/2}},~~~~~~~~z\gg 1.
\end{equation}
The Zeeman part of the vector potential  $A_Z(r)$ is determined by the product of these functions and diminishes  at $r\gg \lambda$ as $\propto 1/r^3$.

On the other hand at  small distances  these functions behave as
\begin{equation}
K_1(z)\approx\frac{1}{z},~~~~~~~~~~~~~z\ll 1,
\label{K}
\end{equation}
and
\begin{equation}
C(z)\approx\frac{4\pi\varepsilon\Delta_\infty^2\xi_v^2}{\lambda}\ln\frac{\lambda z}{\xi_v}, ~~~~~~~~~~~~z\ll 1.
\label{C}
\end{equation}
For the superfluid velocity given by
\begin{equation}
v_s(r)=\frac{\phi_0}{2\pi r}\left (\frac{RK_1(R/\lambda)}{\xi_vK_1(\xi_v/\lambda)}- \frac{2\pi }{\phi_0}RK_1(R/\lambda) C(R/\lambda)\right )
\end{equation}
at $r\ll \lambda$ we obtain
\begin{equation}
v_s(r)\approx\frac{\phi_0}{  2\pi r}\left (1-\frac{8\pi^2\varepsilon\Delta_\infty^2\xi_v^2}{\phi_0}\ln\frac{R}{\xi_v} \right ).
\end{equation}
The  magnitude of dimensionless combination $\frac{8\pi^2\varepsilon\Delta_\infty^2\xi_v^2}{\phi_0}$ is estimated as follows.
Using the relations $\Delta_\infty^2\xi_v^2\approx v_F^2$ and  $\varepsilon=N_0\mu \Im\mathfrak{m} \Psi^\prime(w)/2\pi T\approx7\zeta(3)m^*k_F\mu^2B/(2\pi^2)^2T^2$ (we remind that $ w=\frac{1}{2}-\frac{i\mu B}{2\pi T}$) 
we come to
\begin{equation}
\frac{8\pi^2\varepsilon\Delta_\infty^2\xi_v^2}{\phi_0}=\frac{7\zeta(3)}{\pi^3}k_Fr_e\frac{\varepsilon_F}{T}\frac{\mu B}{T}\ll 1.
\label{estimation}
\end{equation}
Here $k_F$ is the Fermi momentum, $r_e=e^2/mc^2$ is the classical radius of the electron,  the product
of these values is
$k_Fr_e\approx 10^{-5}$. The ratio of the Fermi energy in heavy fermionic compounds to the temperature $\varepsilon_F/T$ taken at temperatures $T\sim T_c/2$   is not larger than $10^2\div10^3$. The ratio $\mu B/T\approx\mu B/T_c$ is smaller than 1 even in the compounds with heavy electron mass. Thus, inequality (\ref{estimation}) is certainly carried out.

Hence, we come to the usual expression for the superfluid velocity
\begin{equation}
v_s(r)\approx\frac{\phi_0}{2\pi r}
\label{velocity}
\end{equation}
which is valid at $r\ll\lambda$. Furthermore, as far as we are working at $B\gg H_{c1}$ the Zeeman term $\varepsilon({\bf h}_v\hat z-B)\Delta$ yields just negligibly small correction
to the  regular term $\alpha(B)\Delta$ in the GL equation. It  means that  the solution of the GL equation for the isolated vortex has  the coordinate dependence
$$
\frac{|\Delta(r)|}{\Delta_\infty}\sim\left\{\begin{array}{rcl} \frac{r}{\xi},~~~~r\ll  \xi \\
~~~~~~~~~~~~~~~~~~~\\
1-\frac{\xi^2}{r^2},~~~~r\gg \xi\end{array}\right.
$$
as it is in the absence of the Zeeman interaction. \cite{Abrikosov} Here, $\xi=\sqrt{\gamma/|\alpha|}$ is the coherence length. This justifies  the variational approach making use the order parameter given by Eqs. (\ref{OP}), (\ref{f}).

The total field amplitude $h_{orb}(r)+h_{Z}(r)$ differs significantly from its orbital part $h_{orb}(r)$ only at the distances of the order of $\xi_v$ from the vortex axis (see Fig. 1).
Hence, the energy of a single vortex is  dominated by the orbital field $h_{orb}(r)$.
Therefore the minimization of the energy of a single vortex gives the same variational parameter $\xi_v=\sqrt{2}\,\xi$ as exposed in the Clem paper.\cite{Clem}
The lower critical field  keeps the usual value determined with logarithmic accuracy as  $H_{c1}\approx\frac{\phi_0^2}{4\pi\lambda^2}\ln\kappa$.

In the high-$\kappa$ limit, we can find a simpler expression to the Zeeman internal field by looking at its behavior at distances
$r\ll \lambda$. By using Eqs. (\ref{K})
and (\ref{C}) we find the dominating term in Eq. (\ref{heps})
\begin{equation}
 h_{Z}\simeq4\pi\varepsilon\Delta_\infty^2\frac{\xi_v^2}{R^2}.
\label{hepseq}
\end{equation} 
We see that $h_Z(r)$ has significant variations near the core of a vortex, the characteristic length associated to it being of the order of $\xi_v$ (see Fig.1). Also the local field amplitude increases with increasing field because $\varepsilon$ is proportional to $B$. The formula (\ref{hepseq}) may be derived by considering the equation
$
 \nabla\times\mathbf{h}_Z=4\pi{\bf j}_Z,
$
that is valid in the absence of the orbital current.  
From Eq. (\ref{heps}) one can find the correction to Eq. (\ref{hepseq}),
$
\delta h_Z=-4\pi \varepsilon\Delta_\infty^2K_0(R/\lambda)\ln(R/\xi_v)/\kappa^2$.

The fast variations of local magnetic field given by the Zeeman interaction in the vicinity of the vortex axis drastically changes the form factor field dependence.
One can calculate the Fourier transform of the  the Zeeman internal field around a single vortex 
\begin{equation}
{h}_Z(q)= \int \ud^2\mathbf{r}\,h_Z(r)e^{-i{\bf q}\cdot{\bf r}}=2\pi\int_0^\infty rdrJ_0(qr)h_Z(r)
\label{Fourier}
\end{equation}
using  Eq. (\ref{hepseq}) valid for $r\ll \lambda$. Indeed, one can prove that the correction to the Fourier transform brought by the region $r>\lambda$ is negligibly small if $q^{-1}\ll\lambda$ (that is when the distance between the vortices is much smaller than the penetration depth).  Furthermore the accuracy of the approximation (\ref{hepseq}) was validated by numerical integration of the full expression (\ref{heps}).
Therefore in what follows we can work with expression (\ref{hepseq}) instead of the more cumbersome Eq. (\ref{heps}), the developed approach is applicable to the  description  of the VL under the fields which are much larger than the lower critical field and not too close to the upper critical field.

By using the coefficients of the Ginzburg-Landau decomposition (\ref{GL}), we obtain the full expression for the coherence length
\begin{equation}
 \xi^2=\frac{v_F^2\langle|\psi(\mathbf{\hat{k}})|^2\hat k_x^2\rangle\Re\mathfrak{e}\Psi^{(2)}(w)}
 {8(2\pi  T)^2\big [\ln(T/T_c)+\Re\mathfrak{e}\Psi(w)-\Psi (1/2)\big ]}.
\end{equation}
For the fields far enough from the critical field this formula gives a slight decrease of the coherence length with increasing field before yielding a strong coherence length increase below the critical line. 

The contribution to the form factor that originates from the interaction of the electron spins with the local magnetic field for an array of $B/\phi_0$ vortices per $cm^2$ is
\begin{equation}
F_{Z}=\frac{B}{\phi_0}{h}_Z(q)=\frac{8\pi^2\varepsilon \Delta_\infty^2
\xi_v^2}{\phi_0}BK_0(q\xi_v).
\end{equation}
After substitution of the explicit expressions of all the values it is
\begin{equation}
 F_{Z}=\frac{4\pi N_0v_F^2\mu B}{\phi_0{T}}\frac{\langle|\psi(\mathbf{\hat{k}})|^2\hat k_x^2\rangle}{\langle|\psi(\mathbf{\hat{k}})|^4\rangle} \Im\mathfrak{m}\Psi'(w)K_0(\sqrt{2}q\xi),
 \label{ZFF}
\end{equation}
where we used the relation $\Delta_\infty^2\xi^2=v_F^2\langle|\psi(\mathbf{\hat{k}})|^2\hat k_x^2\rangle/2\langle|\psi(\mathbf{\hat{k}})|^4\rangle$. The gap averages corresponding to the s- and d-wave cases around a cylindrical Fermi surface are pointed out in the beginning of the Section and show that the form factor is bigger in the s-wave case than in the d-wave case by a factor of 3/2.

Let us also rewrite the form factor orbital contribution given by Eq. (\ref{Fgamma})  with the coefficients of the model
\begin{equation}
 F_{orb}=\frac{4\pi v_F^2}{\phi_0}\frac{\langle|\psi(\mathbf{\hat{k}})|^2\hat k_x^2\rangle}{\langle|\psi(\mathbf{\hat{k}})|^4\rangle}|\alpha|\sqrt{2}q\xi K_1(\sqrt{2}q\xi).
 \label{Fgamma'}
\end{equation}
For $\mu B<2\pi T$ we have $\Im\mathfrak{m}\Psi'(w)\simeq7\zeta(3)\mu B/\pi T$ and $|\alpha|\simeq N_0\big(\ln(T_c/T)-7\zeta(3)(\mu B/2\pi T)^2\big)$. The ratio of the form factors (\ref{ZFF}) and (\ref{Fgamma'}) becomes
 \begin{eqnarray}
 \label{Ratio}
\frac{F_Z}{F_{orb}}\simeq ~~~~~~~~~~~~~~~~~~~~~~~~~~~~~~~\\ 
\frac{7\zeta(3)\mu^2 B^2}{\pi T^2}\frac{K_0(\sqrt{2}q\xi)}{\sqrt{2}q\xi K_1(\sqrt{2}q\xi)[\ln(T_c/T)-7\zeta(3)(\mu B/2\pi T)^2]},\nonumber
\end{eqnarray}
here $\zeta(x)$ is the Riemann zeta function. For $\sqrt{2}q\xi\sim1$ we observe that the Zeeman part of the form factor prevails over its orbital part in the phase diagram region where $\mu B \simeq T$. 
It becomes clear why such a type of behavior occurs only in the superconductors
with dominant paramagnetic depairing effect. Indeed at temperatures $T\sim T_c/2$ one can estimate $\mu B/T \sim B/H_p$. 
For a superconductor characterized by a small Maki parameter  the magnetic field $B$ does not exceed the orbital critical field $B\sim H_{c20}$, therefore the effect of paramagnetism becomes unobservable.
The ratio of the form factors depends from Fermi velocity only through the coherence length. The dominant role played by the Zeeman part of the form reveals itself at small enough coherence length  which corresponds to the sufficiently small Fermi velocity.\\
\begin{figure}[b]
\centering
\includegraphics[width=7cm]{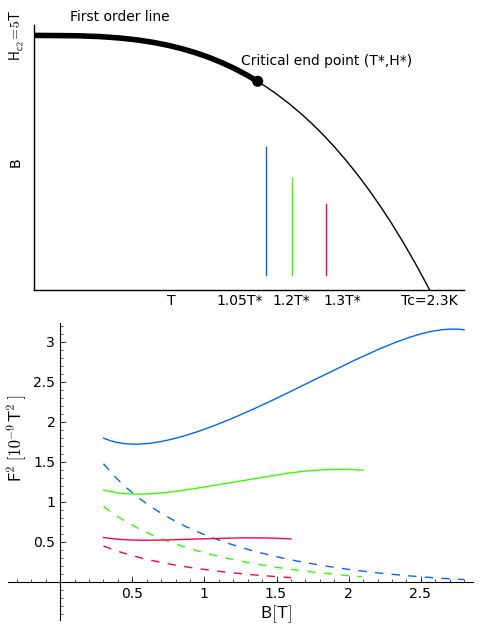}
\caption{(Above) CeCoIn$_5$ phase diagram for $B||$c-axis. The color lines represent the temperatures at which we applied the model. (Below) Variations of the squared form factor $F^2$ at different temperatures including both the orbital and Zeeman contributions. The dashed lines represent the variations of the orbital part only.}
\label{FormFactor}
\end{figure}

We have plotted in Fig. \ref{FormFactor} the total form factor 
\begin{equation}
F=F_{orb}+F_Z, 
\label{sumFF}
\end{equation}
where  $F_{orb}$ and $F_Z$ are given by Eqs. (\ref{CFF}) and (\ref{ZFF}) correspondingly. In numerical calculations we assumed the values $\mu=g\mu_B/2=\mu_B$ for the electron magnetic moment in the material, and $v_F=5\times 10^5\,cm/s$ for the Fermi velocity inside the superconducting phase. A value for $v_F$ slightly bigger was given in Ref. 10 as a result of measurements of the upper critical field $H_{c2}$ near $T_c$. The 2-D density of states on the Fermi surface
is given by $N_0=m^*/2\pi\ell_c$, where we considered $m^*=100\,m_e$ for the electron effective mass, and $\ell_c=7.6\times10^{-8}\,cm$ is the lattice c-axis spacing. In the form factor variations, there is first a domination of the orbital part in the low magnetic field region ($F_Z$ vanishes at $B=0$). We observe next a crossover to a region where the paramagnetic term is dominant. The regime where (\ref{CFF}) goes exactly against (\ref{ZFF}) is likely to explain the observed constant logarithm of the squared form factor\cite{Esk} in the interval $B=0.5-2T$. In addition, these anomalous form factor variations were observed in the experiments realized on the s-wave superconductor TmNi$_2$B$_2$C.\cite{Esk2}

\section{Vicinity of phase transition}

The observed form factor\cite{Bianchi,White} falls towards zero near the phase transition 
to the normal state. Its field dependence in this region is out of applicability of the derivation that makes use of the assumption of isolated noninteracting vortices. The form factor 
decrease at the approach of the upper critical field can be found in the frame of
the GL theory valid in the vicinity of the transition at temperatures above the tricritical point. 

In vicinity of the phase transition  the local magnetic field \cite{Hou2} deviates from its average value $B=\overline{h(\mathbf{r})}$ by
\begin{equation}
 h_1(x,y)=-4\pi\varepsilon(|\Delta(x,y)|^2-\overline{|\Delta|^2}).
 \label{h_1}
\end{equation}
The order parameter solves the linearized Ginzburg-Landau equation. For a square vortex lattice with 
period $a=\sqrt{\pi/eB}$ it  is
\begin{equation}
 \Delta(x,y)=C\sum_{n=-\infty}^{+\infty}(-1)^n e^{2\pi iny/a}e^{-\frac{1}{2\lambda_B^2}(x-na+a/2)^2},
\end{equation}
so that $|\Delta(ma,na)|=0$ (m and n are integer). Here, $\lambda_B=1/\sqrt{2eB}$ is the magnetic length. Multiplying the expression 
by its complex conjugate we have
\begin{eqnarray}
 \nonumber|\Delta(x,y)|^2&=&C^2\sum_{m,n=-\infty}^{+\infty}(-1)^{n+m}e^{-\frac{1}{2\lambda_B^2}(x-na+a/2)^2}\\&&\times e^{-\frac{1}{2\lambda_B^2}(x-ma+a/2)^2} e^{2\pi i(n-m)y/a}.
\end{eqnarray}
Now, putting the dummy index $n'=n-m$, and using Poisson's summation formula\cite{Abrikosov} in the form 
$$\sum_{m=-\infty}^{+\infty}f(x-ma)=\frac{1}{a}\sum_{m=-\infty}^{+\infty}\tilde{f}(\frac{2\pi m}{a})e^{2\pi i mx/a},$$
 where 
 $$
 \tilde{f}(q)=\int_{-\infty}^{+\infty}\mathrm{d}x\,f(x)e^{-iqx}
 $$
 is the Fourier transform of $f(x)$,
we get the quantity of interest
\begin{eqnarray}
 |\Delta(x,y)|^2=~~~~~~~~~~~~~~~~~~~\nonumber\\
 \frac{C^2}{\sqrt{2}}\sum_{m,n=-\infty}^{+\infty}(-1)^{m+n+mn} e^{-\frac{\pi}{2}(m^2+n^2)}
 e^{2\pi i(mx+ny)/a}.
 \label{D^2}
\end{eqnarray}
Its average value is
\begin{equation}
 \overline{|\Delta(x,y)|^2}=\frac{C^2}{\sqrt{2}}.
 \label{barD^2}
\end{equation}
Taking into account Eqs. (\ref{D^2}) and (\ref{barD^2}), the Fourier transform of the magnetic field (\ref{h_1}) yields the form factor corresponding to the Bragg peak with indices (m,n) 
\begin{equation}
 F_{mn}=4\pi\varepsilon\overline{|\Delta|^2}(-1)^{m+n+mn}e^{-\frac{\pi}{2}(m^2+n^2)}.
\label{AFF}
\end{equation}
The field dependence of $\overline{|\Delta|^2}$ at $B\sim H_{c2}(T)$ is known. \cite{Hou2}
We write it in the limit of large GL parameter $\kappa\gg1$
\begin{equation}
\overline{|\Delta|^2}\simeq\frac{\varepsilon(H_{c2}-B)}{2\beta_A\beta}.
\end{equation}
Here $\varepsilon=\varepsilon(H_{c2})$ and $\beta_A=\overline{|\Delta|^4}/\overline{|\Delta|^2}^2$ is the Abrikosov parameter. The critical field 
$$H_{c2}(T)=B_{c}(T)(1-2e\gamma/\varepsilon)
$$
is somewhat lower than $B_c(T)$ determined by the equation $\alpha(B,T)=0$.\cite{Hou2} 
Finally we obtain for the form factor near the phase transition line 
\begin{equation}
 F=|F_{01}|=\frac{2\pi e^{-\frac{\pi}{2}}\varepsilon^2(H_{c2}-B)}{\beta_A\beta}.
 \label{AFFF}
\end{equation}

When the transition becomes first order but near the tricritical point the form factor keeps the form of the one calculated and given by Eq. (\ref{AFF}). However, the main difference here is that the order parameter takes on a finite value at the transition. Hence, the form factor discontinuously falls down to zero at the transition to the normal state.

\section{Conclusion}

Making use of the generalized Clem approach we have calculated the magnetic field dependence of the vortex lattice form factor. The interaction of the electron spins with the space inhomogeneous magnetic field existing inside the superconductor in the mixed state produces diamagnetic currents of pure paramagnetic origin.\cite{Hou2} 
The corresponding field concentrated around the vortex cores yields a dominant contribution to the expression of the vortex lattice form factor at high magnetic field in the superconductors with a small enough coherence length (that was related here to the Fermi velocity). Finally, our analysis showed that the form factor behaves as the square of the order parameter in the region close to the upper critical field and therefore it falls down to zero at the transition to normal state. The fall must be discontinuous when the transition becomes first order. These results give a physical mechanism for the occurrence of anomalous VL form factor variations and possibly account for the measurements that were recently made on CeCoIn$_5$.

\acknowledgments

This work was partly supported by the grant SINUS of the Agence Nationale de la Recherche.

\end{document}